\documentclass[prl, twocolumn, aps,amsmath,amssymb,amsfonts, superscriptaddress]{revtex4-1}

\usepackage{graphicx}
\usepackage{graphics}
\usepackage{amsmath}
\usepackage{amssymb}
\usepackage{amsfonts}
\usepackage{amssymb}
\usepackage{epstopdf}
\usepackage{makeidx}
\usepackage{epsfig}
\usepackage{amsfonts}
\usepackage{bm}
\usepackage{color}
\usepackage{xcolor}
\usepackage{bbold}
\usepackage{afterpage}
\usepackage[colorlinks,allcolors=blue]{hyperref}
\usepackage{booktabs}

\begin{document}
\title{Quantum metric quadrupoles in elemental bismuth thin films}

\author{Rhonald~Burgos~Atencia}
\thanks{These authors contributed equally to this work}
\affiliation{Dipartimento di Fisica ``E. R. Caianiello", Universit\`a di Salerno, IT-84084 Fisciano (SA), Italy}

\author{Pavlo~Makushko}
\thanks{These authors contributed equally to this work}
\affiliation{Helmholtz-Zentrum Dresden-Rossendorf e.V., Institute of Ion Beam Physics and Materials Research, 01328 Dresden, Germany}

\author{Gabriele Naselli}
\altaffiliation{Present address: Department of Physics, Maynooth University, Maynooth, Co. Kildare, Ireland}
\affiliation{Dipartimento di Fisica ``E. R. Caianiello", Universit\`a di Salerno, IT-84084 Fisciano (SA), Italy}
\affiliation{Institute for Theoretical Solid State Physics, IFW Dresden and W\"{u}rzburg-Dresden Cluster of Excellence ct.qmat, Helmholtzstr. 20, 01069 Dresden, Germany}

\author{Debottam~Mandal}
\affiliation{Dipartimento di Fisica ``E. R. Caianiello", Universit\`a di Salerno, IT-84084 Fisciano (SA), Italy}

\author{Paul~Chekhonin}
\affiliation{Helmholtz-Zentrum Dresden-Rossendorf e.V., Institute of Ion Beam Physics and Materials Research, 01328 Dresden, Germany}
\affiliation{Helmholtz-Zentrum Dresden-Rossendorf e.V., Institute of Resource Ecology, 01328 Dresden, Germany}

\author{Sergey~Kovalev}
\affiliation{Department of Physics, TU Dortmund University, 44227 Dortmund, Germany}

\author{Steffen~Kober}
\affiliation{Department of Physics, TU Dortmund University, 44227 Dortmund, Germany}

\author{Zhe~Wang}
\affiliation{Department of Physics, TU Dortmund University, 44227 Dortmund, Germany}

\author{Igor~Veremchuk}
\affiliation{Helmholtz-Zentrum Dresden-Rossendorf e.V., Institute of Ion Beam Physics and Materials Research, 01328 Dresden, Germany}

\author{Oleksiy~Pashkin}
\affiliation{Helmholtz-Zentrum Dresden-Rossendorf e.V., Institute of Ion Beam Physics and Materials Research, 01328 Dresden, Germany}

\author{Fabian~Ganss}
\affiliation{Helmholtz-Zentrum Dresden-Rossendorf e.V., Institute of Ion Beam Physics and Materials Research, 01328 Dresden, Germany}

\author{Maria~Teresa~Mercaldo}
\affiliation{Dipartimento di Fisica ``E. R. Caianiello", Universit\`a di Salerno, IT-84084 Fisciano (SA), Italy}

\author{Denys~Makarov}
\email{d.makarov@hzdr.de}
\affiliation{Helmholtz-Zentrum Dresden-Rossendorf e.V., Institute of Ion Beam Physics and Materials Research, 01328 Dresden, Germany}

\author{Carmine~Ortix}
\email{cortix@unisa.it}
\affiliation{Dipartimento di Fisica ``E. R. Caianiello", Universit\`a di Salerno, IT-84084 Fisciano (SA), Italy}

\maketitle

\textbf{
The nonlinear transport properties of solids are deeply rooted in the quantum geometry of their electronic wavefunctions, which is encoded in the quantum geometric tensor. 
Its real part, known as the quantum metric, has been recently identified as a primary origin of nonlinear transport in quantum materials where time-reversal and inversion symmetries are not simultaneously present. Consequently, the influence of the quantum metric on the largest class of materials -- non-magnetic and centrosymmetric systems -- has remained entirely elusive.
Here, we demonstrate that third-order transport in centrosymmetric materials hosting relativistic fermions is governed by quantum metric quadrupoles (QMQs). We show that these QMQs can originate from both the non-Abelian quantum geometry of bulk three-dimensional Dirac fermions and the Abelian quantum geometry of spin-orbit-coupled surface states.
In stark contrast to all zero-field nonlinear transport signatures known to date, the current driven by these QMQs persists as a robust, non-vanishing observable even in highly scalable polycrystalline thin films. We experimentally validate this quantum metric footprint by measuring nonlinear transport in thin films of elemental bismuth, observing a robust, surface-dominated, and broadband third-harmonic generation that persists up to room temperature. Our findings uncover a hidden role of the quantum metric in polycrystalline systems, establishing third-order nonlinear transport as a high-precision 
diagnostic tool
of wavefunction geometry under ambient conditions.}

\begin{figure*}[tbp]
\includegraphics[width=0.8\linewidth]{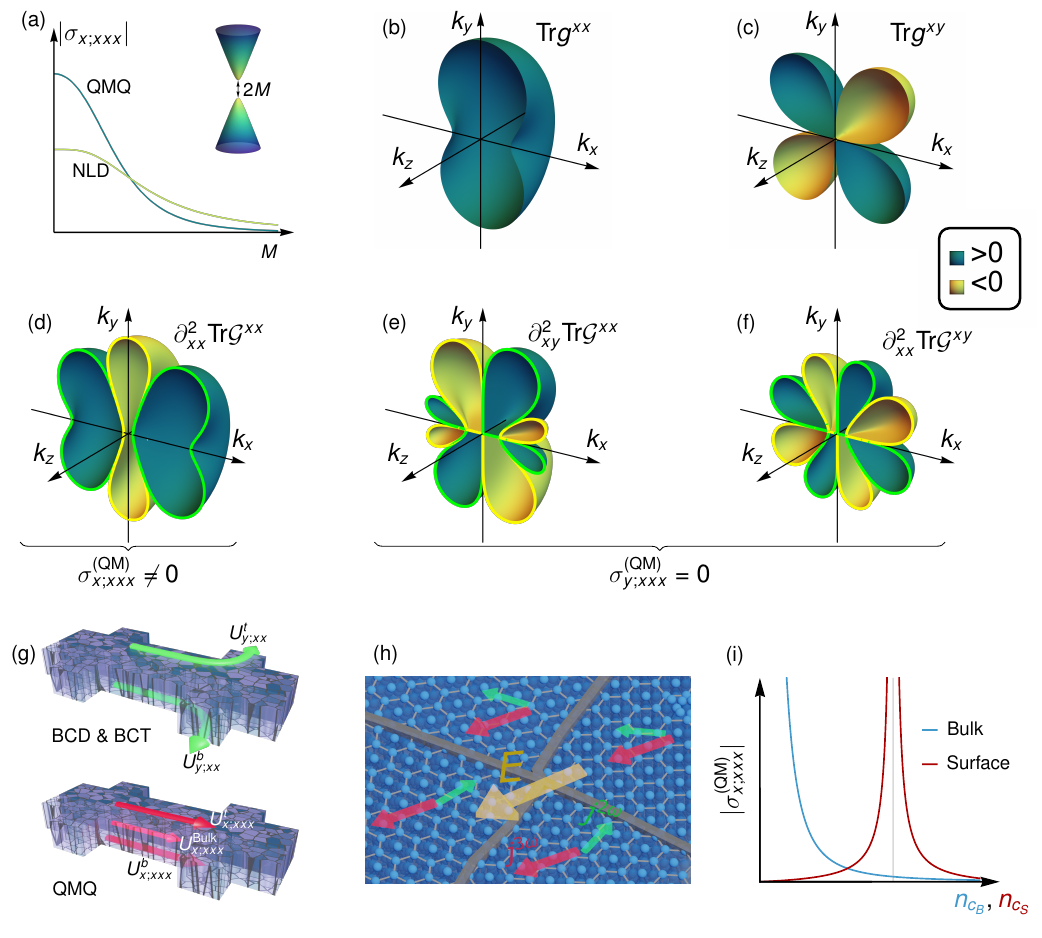}
\caption{
{\bf Quantum metric and nonlinear transport in non-magnetic materials with centrosymmetry} (a) Comparison between the quantum metric quadrupole (QMQ) and nonlinear Drude (NLD) contribution to the third-order longitudinal conductivity  $\sigma_{x;xxx}$  for bulk three-dimensional Dirac fermions with mass $M$. (b)-(c) Three-dimensional contour plots of the trace of the non-Abelian quantum metric $g$  for an isotropic bulk Dirac model. The two quantum metric components $xx$ (b) and $xy$ (c) involved in nonlinear transport are shown.  The 3D contour plots show only representative isosurfaces. 
(d)-(f) 3D-contour plots of the QMQ densities (corresponding to the second derivatives of the trace of the non-Abelian band-energy normalized quantum metric ${\cal G}$) that govern the nonlinear longitudinal $\sigma_{x;xxx}$ and the transversal $\sigma_{x;yyy}$ conductivities. The ensuing  QMQs are found by integration over the Fermi sphere. We also highlight the countours at $k_z=0$ which correspond to the QMQ densities of spin-orbit coupled surface states. (g) Qualitative difference between the second-harmonic currents driven by Berry curvature dipoles and triples and the third-harmonic currents in non-magnetic materials  with centrosymmetric crystalline arrangements. Second-harmonic currents can only flow at crystal terminations and cancel each other at opposite surfaces. Third-harmonic surface currents travel in the same direction and coexist with the symmetry-allowed bulk contribution. (h) In polycrystalline microstructures second-harmonic surface currents suffer from domain averaging and cancel in extended samples. Third-harmonic currents instead have an average component collinear with the driving electric field. (i) Schematic comparison between the bulk quantum-geometry-induced longitudinal third-order conductivity of bulk massless Dirac fermions that decreases monotonically with $n$-type carrier doping and the surface third-order conductivity displaying instead a non-monotonous behavior with a divergence at the Lifshitz transition.
} 
\label{fig:NAQM}
\end{figure*}

The electronic wavefunctions of a solid define a gauge-invariant quantum geometric tensor~\cite{pro80} (QGT) whose imaginary part constitutes a field known as Berry curvature (BC). The flux of the BC is finite in magnetic materials where it manifests 
itself
as 
an
intrinsic component of the anomalous Hall conductivity \cite{Haldane2004,Nagaosa2010,Sinitsyn2008}. 
Beyond the linear regime, electronic transport can carry signatures of higher-order moments of the BC in time-reversal invariant materials with non-centrosymmetric crystalline structures. Specifically, the second-order nonlinear Hall effect with time-reversal symmetry is governed by BC dipoles (BCDs)~\cite{Sodemann2015,Ortix2021,Du21,Suarez2025,ma19,les23,kan19,ho21} and triples (BCTs)~\cite{iso20,he21,mak24}. 

While these phenomena originate from the imaginary part of the QGT, the physical implications of its real part -- the quantum metric quantifying the distance between electronic wavefunctions in crystalline momentum space --  have only recently begun to be explored~\cite{Torma_2023,Yu_2025}. The quantum metric enables flows of supercurrent in flat bands~\cite{Peotta_2015} and contributes significantly to electron-phonon coupling~\cite{Yu_2024}. Furthermore, in systems with broken time-reversal symmetry the quantum metric yields an intrinsic contribution to 
the
second-order nonlinear electronic transport~\cite{gao14,das23,kap24,Jiang2025}, as experimentally observed in antiferromagnetic topological insulators~\cite{gao23,wan23exp}, 
and
via nonlinear magnetoresistance at oxide interfaces~\cite{Sala_24} and on the surfaces of three-dimensional topological insulators~\cite{Mercaldo2025,Sala_2026}.

Unlike the Berry curvature, the simultaneous presence of time-reversal and bulk inversion symmetries does not force the quantum metric to vanish, allowing it to remain finite both in the bulk and at the material surfaces. Here, we demonstrate that these quantum metric properties can be directly probed via third-order nonlinear transport. Contrary to other zero-field transport signatures of quantum geometry -- whose observation is 
restricted to high-quality single crystals -- the third-order nonlinear conductivity tensor possesses an isotropic component that survives spatial averaging over domains with random crystallographic orientations. Consequently, the quantum-metric-induced currents 
persist as a robust, macroscopic transport signature even in highly scalable and technologically relevant polycrystalline thin films. Third-order responses thus emerge as a definitive diagnostic tool to unveil the quantum geometric properties of the most abundant class of materials: non-magnetic systems with inversion symmetry.

We 
predict that materials with linear bulk or surface electronic bands generally feature a substantial  third-order nonlinear transport response induced by quantum metric quadrupoles (QMQ)~\cite{Fang_2024,Mandal_2024}. A wide range of material platforms, spanning from Dirac semimetals such as 
BiO$_2$~\cite{Young_2012} and Cd$_3$As$_2$~\cite{Armitage_2018} to conventional metals with Rashba surface states including Au~\cite{Yan_2015} 
are thus expected to host a sizable, geometry-induced third-order nonlinear transport when fabricated in thin-film form. Importantly,  QMQs can also drive optical third-harmonic generation (THG) -- a key capability in modern electronics that enables the upconversion of electronic signals to much higher frequencies --  in the technologically relevant terahertz (THz) spectral range~\cite{Salikhov_2025}. This quantum geometry-driven  
THG can be highly efficient, largely exceeding surface-induced second-order nonlinearities, which inherently suffer from domain averaging and signal cancellation in polycrystalline systems. We experimentally validate these theoretical predictions in polycrystalline thin films of elemental bismuth, providing direct evidence of a surface-dominated, quantum metric-induced third-order transport at room temperature, accompanied by a highly efficient THz THG.

Treating the dynamics of Bloch electrons at the semiclassical level shows that the third-order nonlinear conductivity tensor defined by the relation $j_{\alpha}=\sigma_{\alpha;\beta \gamma \delta} E_{\beta} E_{\gamma} E_{\delta}$ contains two fundamentally distinct contributions. The first 
is a geometric contribution, which, due to the enforced twofold degeneracy of the electronic bands, is linked to a non-Abelian quantum metric $g$~\cite{Ma_2010,Ding_2024}.
Specifically, this third-order conductivity is governed by the trace of the non-Abelian band-energy normalized quantum metric (BNQM) ${\mathcal G}$ [see the Supplemental Material]. This geometric term scales linearly with the relaxation time $\tau$ differing from the second, nonlinear Drude (NLD) contribution, which depends exclusively on the group velocity of the Bloch electrons and scales with $\tau^3$. 
To evaluate this effect in a realistic setting, we consider the minimal bulk electronic model hosting a finite non-Abelian QGT: a four band Dirac isotropic model described by the Hamiltonian ${\mathcal H}={\bf d} \cdot {\boldsymbol \Gamma}$ with ${\boldsymbol \Gamma}$ the five-dimensional vector of the anticommuting Dirac matrices. For this system, we find [see the Supplemental Material and Fig.~\ref{fig:NAQM}(a)] that the QMQ contribution largely dominates over the NLD contribution in the small Dirac gap regime. Additionally, the traces of the non-Abelian quantum metric [see Fig.~\ref{fig:NAQM}(b),(c)] can be directly related to the five-dimensional ${\bf d}$ vector of the Dirac Hamiltonian via the expression $\textrm{Tr}(g^{\mu\nu})=\partial_{k_\mu} \hat{\bf d} \cdot \partial_{k_\nu} \hat{\bf d}/2$. By additionally using that the BNQM trace  $\textrm{Tr}({\cal G}^{\mu \nu})=\textrm{Tr}(g^{\mu\nu}) / |{\bf d}|$, we can directly map the momentum space distribution of the QMQ densities. In particular, the $\partial_{xx}^2 \textrm{Tr} {\cal G}^{xx}$ component, which governs the longitudinal nonlinear conductivity $\sigma_{x;xxx}$, yields a finite net value [see Fig.~\ref{fig:NAQM}(d)]  upon integration over the spherical Fermi surface. Conversely, the QMQ density components $\partial_{xy}^2 \textrm{Tr} {\cal G}^{xx}$, $\partial_{xx}^2 \textrm{Tr} {\cal G}^{xy}$
have a characteristic octupolar profile [Fig.~\ref{fig:NAQM}(e),(f)] in momentum space, which strictly guarantees the vanishing of any third-order nonlinear 
transversal 
response. 

Having established that three-dimensional Dirac fermions host a third-order electronic transport governed by the non-Abelian BNQM, we turn to the responses originating from electronic surface states. Since inversion symmetry is naturally broken at a crystal termination, second-order nonlinear optical and transport processes can be symmetry-allowed, provided there is no evenfold rotation with axis perpendicular to the surface 
~\cite{Gabriele_2025}. 
At crystal terminations 
with 
low crystalline symmetries a surface Berry curvature dipole can be non-vanishing~\cite{Wawrzik_2023}. Alternatively, a disorder-induced nonlinear Hall effect has been instead observed at the (111) surfaces of
Bi$_2$Se$_3$~\cite{he21}, Bi$_2$Te$_3$
~\cite{Wang_2026} 
and in elemental bismuth thin films~\cite{mak24}, 
where nonlinear skew scattering and side-jump processes are governed by a Berry curvature triple. 
A characteristic fingerprint of the transverse responses induced by these surface Berry curvature dipoles and triples is that the second-order currents at opposite surfaces travel in opposite directions [see Fig.~\ref{fig:NAQM}(g)]. Consequently, a net macroscopic second-order response can only occur if the opposite surfaces are strongly inequivalent. In stark contrast, the longitudinal third-order surface currents -- originating from the Abelian counterpart of the bulk BNQM --  flow in the same direction at opposite surfaces [Fig.~\ref{fig:NAQM}(g)]. This implies that the effect of the bulk QMQs is constructively boosted by the surface  QMQs.

In addition, while the second-order surface currents inherently suffer from domain averaging and signal cancellation in polycrystalline films [see Supplemental Material and Fig.~\ref{fig:NAQM}(h)], 
third-order currents retain a net, non-vanishing average component. As a result, the third-order response driven by bulk  QMQs inevitably coexists with, and is reinforced by, the  QMQ contribution from the surface states.
The QMQ densities characterizing conventional Rashba  surface states share the same momentum dependent characteristics as their bulk  counterparts [see the Supplemental Material and Fig.~\ref{fig:NAQM}(d),(e),(f)]. 
The behavior of the net QMQ while changing the carrier density, instead, is qualitatively different. The total bulk  QMQ of massless Dirac fermions decreases monotonically when moving away from the Dirac point. In stark contrast, the net QMQ of Rashba surface states exhibits a non-monotonic carrier density dependence, and feature a divergence located precisely at the Lifshitz transition of the Rashba bands [see Fig.~\ref{fig:NAQM}(i)]. This unique property implies that surface states can provide a gigantic renormalization of the overall QMQ bulk response.

\begin{figure*}[tbp]
\includegraphics[width=0.8\linewidth]{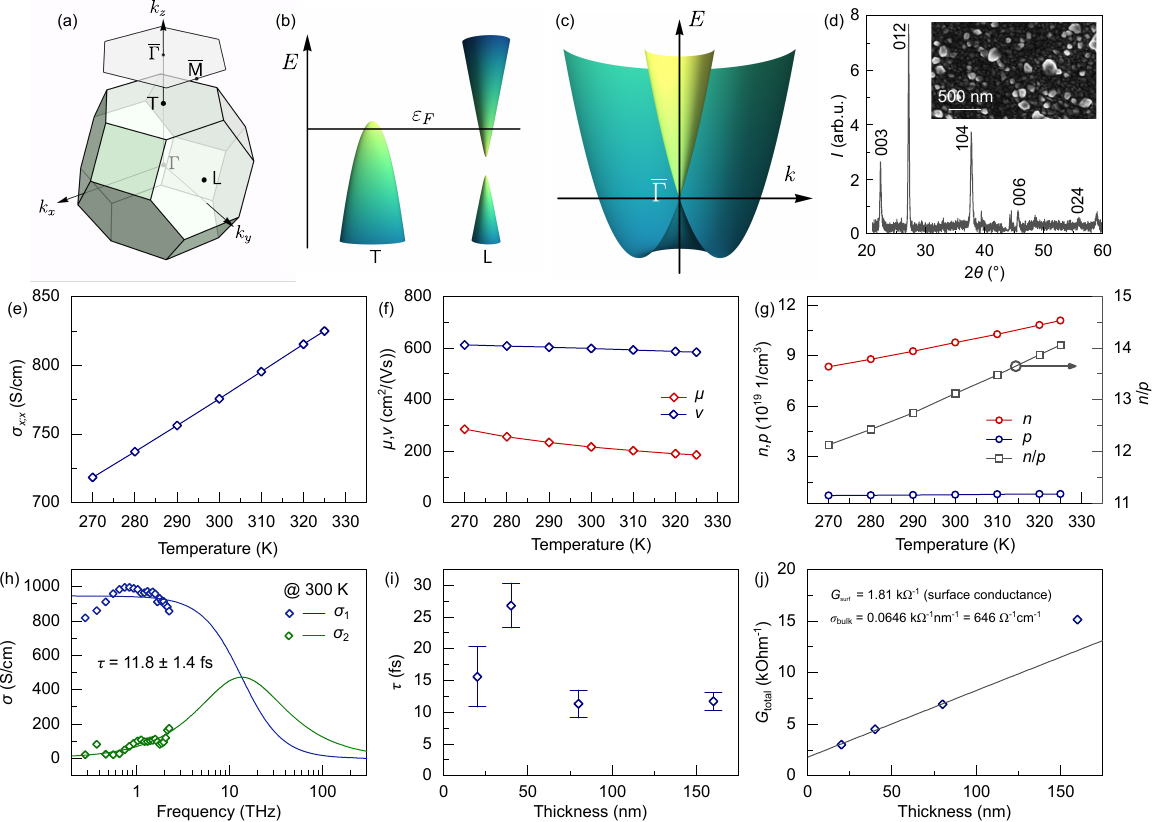}
\caption{\textbf{Electronic properties, magnetotransport and optical transport in Bi thin films.}
(a) 3D Brillouin zone (BZ) of elemental Bismuth. At the T point of the BZ a large hole pocket is found. The three inequivalent $L$ points instead host Dirac electrons with a Dirac mass of few meV. (b) At the $\bar{\Gamma}$ centre of the projected two-dimensional BZ for films grown along the (111) direction, spin-orbit coupled surface states with Rashba spin-splitting appear.
(d) $\theta - 2\theta$ XRD pattern of bismuth thin film deposited onto polyimide substrate. Inset shows high-resolution scanning electron microscopy image of the film`s surface, confirming its polycrystalline nature.
(e) Temperature dependence of linear electrical conductivity $\sigma_{x;x}$ of the bismuth thin film.
(f,g) Temperature evolution of charge carrier (f) mobility and (g) density calculated for bismuth film assuming two-band conductance model.
(h) Optical conductivity of bismuth thin film fabricated on Al$_2$O$_3$ substrate. Symbols represent measured data and lines are the Drude model fit.
(i) Relaxation time and (j) total conductance of the bismuth films of different thickness.
}
\label{fig:LinTrans}
\end{figure*}
Thin films of elemental bismuth represent an ideal candidate platform to observe the coexistence of bulk and surface quantum geometric effects.  
The Fermi surface of elemental bismuth consists of a large hole pocket sitting at the $T$ point of the Brillouin zone (BZ) and three small electron pockets related to each other by a threefold rotation symmetry around the $(111)$ axis [Fig.~\ref{fig:LinTrans}(a)]. This Fermi surface structure defines a valley degree of freedom for the electrons that leads to intriguing transport phenomena at the linear order. These include an high-field magnetoresistance drop due to a complete emptying of one or two valleys above a critical magnetic field~\cite{Zhu_2017}, and an orbital anisotropic magnetoresistance~\cite{Collaudin_2015} with an ensuing planar Hall effect~\cite{Yang_2020}. While the large hole pocket at the $T$ point lacks non-trivial quantum geometric properties, the Dirac nature of the electron pockets combined with their small gap~\cite{Vecchi_1974} in the few meV range  implies the presence of sizable bulk QMQs. In addition, since bismuth is classified as a higher-order topological insulator~\cite{Schindler_2018} with a trivial value of the ${\mathbb Z}_2$ topological invariant~\cite{Fu_2007} for systems in the AII class of the Altland-Zirnbauer classification~\cite{Altland_1997}, it is expected to feature surface states that realize spin-orbit coupled Rashba bands [Fig.~\ref{fig:LinTrans}(b)].

\begin{figure*}
\includegraphics[width=0.95\linewidth]{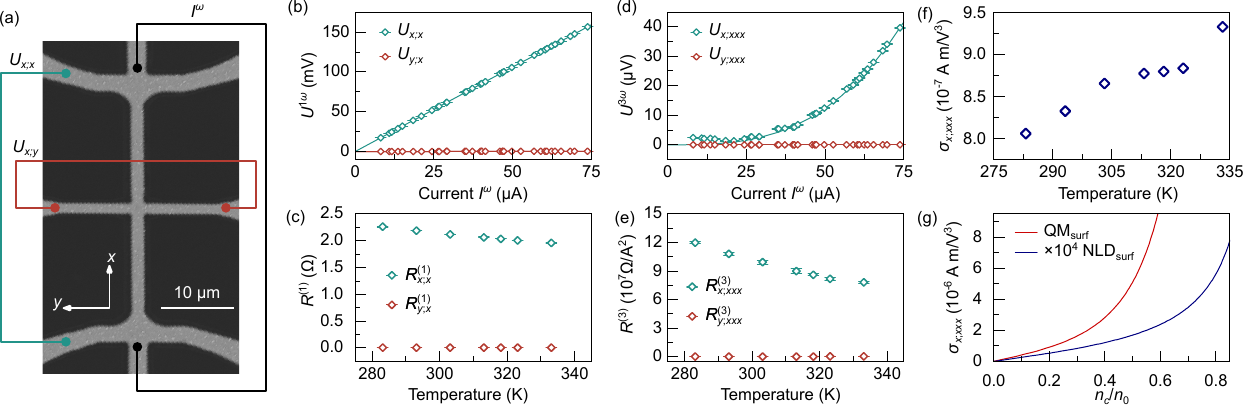}
\caption{\textbf{Nonlinear transport in bismuth thin film on polyimide substrate.}
    (a) Scanning electron microscope image of the Hall bar device. AC electrical current $I^\omega$ is applied along the $x$ direction, longitudinal and transverse voltages are recorded at first, as well as second and third harmonics at temperature of $303$\,K.
    (b) Longitudinal $U_{x;x}$ and transverse $I_{y;x}$ first harmonic voltages  as function of applied current. The $U_{x;x}$ shows linear dependence on applied current, indicating Ohmic behaviour of the device. Symbols represent the measured data and lines are fit.
    (c) Temperature evolution of longitudinal and transverse resistances $R^{(1)} = U^{1\omega}/I^{\omega}$ of the Hall bar device. 
    (d) Longitudinal $U_{x;xxx}$ and transversal $U_{y;xxx}$ third harmonic voltages versus applied current. The longitudinal third harmonic voltage shows cubic dependence on the applied current as $U_{x;xxx} = R^{(3)}_{x;xxx} (I^{\omega})^3$.
    (e) Temperature evolution of $R^{(3)}$ coefficient for longitudinal and transversal third harmonic signal.
    (f) Temperature dependence of third-order longitudinal nonlinear conductivity $\sigma_{x;xxx}$ indicating that nonlinear transport is more favorable with $n$-type carrier doping.
    (g) Theoretical values for the nonlinear longitudinal conductivity for Rashba surface states with effective Hamiltonian ${\mathcal H}=\hbar^2 k^2 / (2 m^{\star}) + \alpha_R (\sigma_x k_y - k_x \sigma_y)$ as a function of the carrier density. We have considered a Rashba spin-orbit strength $\alpha_R=0.5$\,eV\AA, an effective mass $m^{\star}=0.01 m_e$, a relaxation time $\tau=12$\,fs, and a penetration depth of the surface states $l_d=5$~nm. The electronic sheet density is measured in units of $n_0=k_0^2/(4 \pi)$ with the characteristic momentum $k_0=2 m^{\star} \alpha_R / \hbar^2$.
}
\label{fig:NonlinTrans}
\end{figure*}

We fabricated $100$-nm-thick Bi thin films 
by magnetron sputtering. 
Structural characterization using X-ray diffraction  (XRD) , scanning electron microscopy [Fig.~\ref{fig:LinTrans}(c)], and electron backscatter diffraction (EBSD) [see the Supplemental Material] reveal granular microstructure with grain size ranging from 20\,nm up to 150\,nm and preferential rhombohedral $(1\,1\,1)$ crystallographic texture. The thin films are isotropic
within the film plane. Linear transport measurements show a characteristic semiconducting behavior, evidenced by an increase of conductivity with temperature [Fig.~\ref{fig:LinTrans}(e)]. To confirm the simultaneous contribution of both hole and electron carriers to transport, we have performed magnetoresistance and Hall measurements by applying an external out-of-plane magnetic field up to $12$\,T [see Supplemental Material]. 
The nonlinear magnetic field dependence of both the Hall resistance and the magnetoresistance are fitted using a two carrier model \cite{Li2016a} that allow us to extract the temperature dependence of hole and electron densities ($p, n$) and the corresponding mobilities ($\nu, \mu$) [Fig.\,\ref{fig:LinTrans}(f,g)]. At room temperature the electron density is one order of magnitude larger than the holes, reaching values of $n=10^{20}$\,cm$^{-3}$ consistent with previous reports on Bi thin films~\cite{Krupinski2020}. 
The electron mobility $\mu \simeq 2\times10^2$~cm$^2/$Vs is two orders of magnitude smaller than in Bi single crystals~\cite{Michenaud_1972}. 
In single crystals, the effective mass tensor of the electronic ellipsoidal Fermi surfaces is highly anisotropic with the mass along the bisectrix reaching $m^{\star} \simeq 0.2 m_e$, while the masses in the trigonal and binary axes are two orders of magnitude lower~\cite{Liu_1995}.  Consequently, by assuming an averaged isotropic effective mass $\simeq 0.1 m_e$, as appropriate for our polycrystalline samples, we can estimate a short electronic relaxation time $\tau_n \simeq 11$~fs. 
A consistent electronic relaxation time of $\tau = 11.8 \pm 1.4$\,fs was independently extracted by performing optical conductivity measurements fitted with a conventional Drude model [Fig.~\ref{fig:LinTrans}(h)]. Notably, these optical conductivity measurements, systematically carried out across thin films of varying thicknesses [Supplementary Information], demonstrate that the relaxation time is essentially independent of the film thickness [Fig.~\ref{fig:LinTrans}(i)]. Furthermore, by extracting the total sheet conductance as a function of thickness [Fig.~\ref{fig:LinTrans}(j)], we identify a substantial surface conductance $G_\text{surf} \simeq 1.81$~k$\Omega^{-1}$. 
This direct evidence highlights the crucial, cooperative role played by both bulk and surface electronic states in these thin films.

We next address their relevance in the room-temperature  nonlinear transport properties. To this end, we  measured 
the electrical responses of our Bi thin films by sourcing an alternating current while simultaneously recording the first $1 \omega$, second $2 \omega$ [see Supplemental Material], and third $3 \omega$ 
harmonic voltages in both the longitudinal and transverse channels [Fig.~\ref{fig:NonlinTrans}(a)]. 
Fig.~\ref{fig:NonlinTrans}(b) reports the current-voltage characteristic of the first-harmonic longitudinal response, which 
remains strictly linear over the full range of applied currents. This ideal Ohmic behavior confirms that Schottky barriers do not play a significant role in our devices. 
Furthermore, and in agreement with the constraints imposed by the trigonal symmetry of the lattice, 
we find the transverse voltage and its associated resistance to be three orders of magnitude smaller than their longitudinal counterparts 
across the entire temperature range  [Fig.~\ref{fig:NonlinTrans}(c)].  
Fig.~\ref{fig:NonlinTrans}(d) demonstrates the occurrence of a strong third-harmonic longitudinal response. The longitudinal voltage has a characteristic cubic current-voltage dependence, which  
 can be reliably fitted by the relation $U_{x;xxx}=R^{(3)}_{x;xxx} (I^{\omega})^3$ [Fig.~\ref{fig:NonlinTrans}(e)]. 
Independent of temperature, we find that the third-harmonic transverse voltages  [Fig.~\ref{fig:NonlinTrans}(e),(f)] are two orders of magnitude smaller than their longitudinal counterparts.  
Crucially, since the nonlinear conductance $\sigma_{x;xxx}$ [Fig.~\ref{fig:NonlinTrans}(f)] increases with temperature precisely as the concentration of $n$-type carriers, we conclude that the third-harmonic transport of our bismuth thin films is completely dominated by the contribution of the surface states. 
This conclusion is further validated by our theoretical modelling: 
the nonlinear conductivity due to  the Rashba modes [Fig.~\ref{fig:NonlinTrans}(g)] can be theoretically estimated as $\sigma_{x;xxx} \simeq 10^{-6}$~Am/V$^3$ 
assuming a penetration depth of the surface states $l_d=5$~nm. 
This value is remarkably close to our experimental data, and significantly larger than the values expected from the 3D bulk Dirac electrons of Bi [see the Supplemental Material]. 
Finally, the fact that the conductance ratio $\sigma_{x;xxx}/\sigma_{x;x}$ is independent of mobility [see the Supplemental Material] strongly points  toward a purely geometric origin of the measured signal. This is unambigously confirmed by our theoretical estimates, which place the nonlinear Drude contribution of the Rashba surface states  four orders of magnitude below the QMQ term [Fig.~\ref{fig:NonlinTrans}(g)]. 

We have demonstrated, in conclusion, nonlinear third-order transport responses due to quantum metric quadrupoles at room temperature in non-magnetic materials with centrosymmetric crystalline arrangements. Contrary to other nonlinear responses, including the  third-order nonlinear Hall effect observed in T$_d$-MoTe$_2$~\cite{Lai_2021} and Fe$_5$GeTe$_2$~\cite{Yu_2025_THG}, the longitudinal nonlinear conductivity possesses an isotropic component that survives domain averaging and can thus  appear as a macroscopic observable in polycrystalline thin films. Polycrystalline thin films are of technological relevance as they can be integrated with complementary metal oxide-semiconductors. Additionally, as shown by our elemental Bi thin films [see the Supplemental Material] the QMQ-induced nonlinear response can be efficient up to the terahertz range, suggesting the design of upconverters using low-cost microstructures as efficient sources. 

Bismuth thin films display a high third-order conductivity of $\sigma_{x;xxx} \simeq 10^{-6}$A m/V$^3$ 
that is comparable to the transversal conductivity of single crystals. Additionally, elemental bismuth offers other key advantages including the absence of toxic elements, established scalable production, and relevance for electronic applications such as magnetic field sensors~\cite{Melzer_2015}. 
Quantum metric quadrupoles are not limited to bismuth but are expected in all materials with linearly dispersing electronic bands, ranging from Dirac semimetals to noble metals featuring strongly spin-orbit coupled surface states. Our study will thus enable a range of possibilities to harness zero-field quantum metric effects and inspire technologies based on the geometry of electronic wavefunctions.


\section*{Methods}

\subsection*{Sample fabrication}

Bismuth thin films with nominal thickness of $100$\,nm were prepared using RF magnetron sputtering at room temperature (base pressure: better than $10^{-7}$\,mbar; Ar sputter pressure: $10^{-3}$\,mbar; deposition rate: $0.3$\,nm/s). 
For the nonlinear transport characterization [Fig.~\ref{fig:NonlinTrans}], bismuth thin films were deposited onto polyimide foils of $25$-µm thickness (Kapton, DuPont, USA). The films were structured into Hall bar devices via conventional UV optical lithography and lift-off process. The substrate was pre-baked at $120^\circ$C for $5$\,min and surface cleaned in oxygen plasma for $3$\,min. Image reversal photoresist AZ5214e (MicroChemicals GmbH, Germany) was spin coated at $6000$\,rpm for $30$\,s and soft baked on a hot plate at $100^\circ$C for $90$\,s. The samples were exposed using direct laser writer (DWL66, Heidelberg Instruments, Germany), post-baked at $115^\circ$C for $90$\,s and developed in 1:4 solution of AZ351b developer (MicroChemicals GmbH, Germany) in deionized (DI) water. After thin film deposition, extra material was lifted-off in acetone to reveal the device structure and rinsed with isopropanol and DI water.
Electric contacts to the bismuth Hall bar were realized using silver conductive paint (ACHESON Silver DAG 1415, Plano GmbH, Germany). To prevent oxidation of bismuth surface layers, the devices were covered with GE-varnish (Oxford Instruments, UK) after fabrication.

For high-field magnetotransport measurements, bismuth films were deposited onto Si/SiO$_2$(100\,nm) substrates (Crystal GmbH, Germany). The films were structured into Hall bar devices using UV optical lithography and wet-etching technique. The extended films were spin coated with photoresist AZ1518 (MicroChemicals GmbH, Germany) at $4000$\,rpm for $30$\,s. The Hall bar structure was exposed using direct laser writer (DWL66, Heidelberg Instruments, Germany) and developed in 1:4 solution of AZ351b developer (MicroChemicals GmbH, Germany) in deionized (DI) water. The sample is then etched in 1:10 sodium persulfate solution in DI water (B327, AG TermoPasty, Grzegorz Gasowski, Poland) for $120$\,s. The UV exposure is then repeated to prepare the electric contacts. Contact pads of Cr($10$\,nm)/Au($100$\,nm) were deposited using DC magnetron sputtering and a shadow mask.

For optical conductivity characterizations, complementary samples with nominal thicknesses ranging from $20$\,nm to $160$\,nm were grown on Al$_2$O$_3$ substrates (Crystal GmbH, Germany). The samples were spin coated with photoresist AZ1518 (MicroChemicals GmbH, Germany) at $4000$\,rpm for $30$\,s to protect bismuth from oxidation while retaining transparency necessary for optical measurements.

\subsection*{X-ray diffraction}

XRD studies were carried out using a Rigaku SmartLab $3$\,kW with a parallel beam of {Cu}-$K_{\alpha}$ radiation (wavelength: $1.542$\,\AA). Lattice parameter refinement was performed with SmartLab Studio II. 

\subsection*{Electron microscopy}

Scanning electron microscopy (SEM) imaging and electron backscatter diffraction (EBSD) were performed in a Zeiss NVision 40 SEM equipped with a Schottky field emission electron cathode and a Bruker EBSD system with an e-Flash FS detector.
For the EBSD, the acceleration voltage was set to $20$\,kV, the beam current to about $10$\,nA using a $120\,\mu$m aperture. The EBSD detector resolution was set to $240\,\times\,180$ pixels and the exposure time to least $330$\,ms per frame. An EBSD mapping was done as a rectangular grid of $480\,\times\,360$ points with a step size of $600$\,nm.
The evaluation of the EBSD data was done with the Bruker software ESPRIT 2.6.
The minimum number of indexed bands was set to 7 and the maximum mean angular deviation to $0.9^\circ$. These conditions resulted in an indexation rate of $21$\%.

\subsection{Optical conductivity}

Optical conductivity of Bi films was measured using a home-built time-domain THz spectroscopy setup. It is based on a femtosecond Ti:Sapphire laser amplifier with a pulse repetition rate of 250\,kHz. THz pulses are generated by a photoconductive GaAs emitter and detected using an electro-optic sampling in a $(1\,1\,0)$ ZnTe crystal. 
First, we obtained the complex transmission function of a Bi film on sapphire substrate as
\[
t(\omega) = \frac{E_{\mathrm{film}}(\omega)}{E_{\mathrm{sub}}(\omega)} ,
\]
where $E_{\mathrm{film}}(\omega)$ is the spectrum of the THz pulse transmitted through the Bi film on the substrate and $E_{\mathrm{sub}}(\omega)$ is the spectrum of the THz pulse transmitted through a bare part of the substrate that intentionally was not covered by bismuth.

The complex optical conductivity spectra were calculated using the Tinkham equation for thin films~\cite{Meged2023}:
\[
\sigma_\text{opt}(\omega) = \frac{1+n_{\mathrm{sub}}}{Z_0 d} \left( \frac{1-t(\omega)}{t(\omega)} \right),
\]
where $d$ is the thickness of a thin film, $n_{\mathrm{sub}} = 3.1$ is the refractive index of the sapphire substrate, and $Z_0 = 376.7~\Omega$ is the free-space impedance.

The optical conductivity spectra were fitted using the Drude model:
\[
\sigma_\text{opt}(\omega) = \frac{\sigma_0}{1-i \omega \tau},
\]
where $\tau$ is the scattering time and $\sigma_0$ is the DC conductivity. Figure~SI-7 shows the experimental data and the fitting curves for Bi films of thickness ranging from $20$\,nm to $160$\,nm. The Drude fits reveal the scattering times $\tau$ of $12...27$\,fs. Although the experimental THz spectra possess some artefacts that limit the fit quality and the peak of the Drude response lies outside of the measured spectral range, the uncertainty of the scattering time is rather small. This fact is known from multiple studies of different conducting materials \cite{Ulbricht2011} and it is mainly related to the small number of the model parameters (only $\tau$ and $\sigma_0$). In our case, $\sigma_0$ is mainly defined by the real part of the optical conductivity whereas $\tau$ is set by the increase of the imaginary part with frequency.

\subsection*{Magnetotransport and Sheet resistance}

The magnetotransport measurements were carried out using the Zero-Offset Hall preset of a Tensormeter measurement device (Tensor Instruments, HZDR Innovation GmbH, Germany).
The high magnetic field measurements (up to $12$\,T, Fig.~\ref{fig:LinTrans}) were performed on Bi($100$\,nm) film prepared on Si/SiO$_2$($100$\,nm) substrate in TeslatronPT cryostat (Oxford Instruments, UK) with magnetic field applied perpendicular to the film plane. The charge carrier densities and mobilities are estimated based on the two-band model \cite{Li2016a} as:  
\begin{equation*}
    \begin{aligned}
    \rho_{xx} & = \frac{1}{q}
    \frac{(p \nu + n \mu) + (n \mu \nu^2 + p \nu \mu^2) B^2}
    {(p \nu + n \mu)^2 + \mu^2 \nu^2 (p - n)^2 B^2}
    \\
    \rho_{xy} & = \frac{1}{q}
    \frac{(p \nu^2 - n \mu^2) + \mu^2 \nu^2 (p-n) B^2}
    {(p \nu + n \mu)^2 + \mu^2 \nu^2 (p - n)^2 B^2}
    B,
    \end{aligned}
\end{equation*}
where $q$ is an electron charge, $B$ is the magnetic field, $n$ ($p$) and $\mu$ ($\nu$) are the density and mobility of electrons (holes), respectively.

The magnetotransport measurements on Bi($100$\,nm) film prepared on polyimide foil were performed in a custom build magnetotransport measurement setup with the magnetic field (max $1.5$\,T) applied perpendicular to the film plane. The temperature control was realized using a Peltier element and thermistor PT100 positioned in the vicinity of the sample. 
The charge carriers density was calculated assuming only electron conductivity as $n \, =  \, \dfrac{1}{R_\text{H} d q}$, where $R_\text{H}$ -- Hall resistance and $d = 100$\,nm -- film thickness. The mobility of charge carriers was estimated as $\mu = \frac{|R_\text{H}|}{\rho},$ with $\rho$ being resistivity of bismuth.

The third-order nonlinear conductivity of bismuth thin film is estimated as:
\begin{equation*}
\sigma_{x;xxx} = \sigma_{x;x}\frac{U_{x;xxx}}{U_{x;x}
^3} \frac{l_x^3}{l_x},
\end{equation*}
with $\sigma_{x;x}$ being linear conductivity, $U_{x;x}$ and $U_{x;xxx}$ -- measured first and third harmonic harmonic voltages and $l_{x}$ -- dimension of the Hall bar device along the direction of applied current.

\subsection*{Harmonic transport measurement}

Electrical transport harmonic measurements were performed using Tensormeter RTM2 (Tensor Instruments, HZDR Innovation GmbH, Germany). The sinusoidal current at the fundamental frequency of $37.5$\,Hz was sourced using the internal generator. In and out-of-phase components of the voltage signal were recorded for fundamental as well as second and third harmonics in both longitudinal and transversal channels. For each sourcing current value, the data were averaged over 1\,min integration time. A metal shielding around the sample was used to minimize the influence of external parasitic RF signals. The temperature control was realized using a Peltier element and thermistor PT100 positioned in the vicinity of the sample. 

\subsection*{THz harmonic study}

To study THz harmonic generation, we employed narrowband THz pulses centered at $350$\,GHz. The THz source was realized using a tilted pulse front scheme in a LiNbO$_3$ crystal, driven by $100$\,fs, $800$\,nm laser pulses at a $1$\,kHz repetition rate with a pulse energy of about $3$\,mJ. Two band pass filters centered at $350$\,GHz ($20$\% bandwidth) were used to convert the initially broadband THz emission into narrowband radiation. 
The third-harmonic THz radiation emitted from the sample was characterized by electro-optic sampling in a $2$\,mm thick ZnTe crystal. To enhance the sensitivity of the electro-optic detection to the third harmonic, additional THz band pass filters were placed after the sample. These filters efficiently transmit radiation around $1000$\,GHz while strongly suppressing the fundamental component at $350$\,GHz.

\section*{Data availability}
All data that support the plots within this paper and other ﬁndings of this study are available from the corresponding authors upon reasonable request.

\section*{Acknowledgements}
We thank Conrad Schubert (HZDR) for support with thin film deposition and Prof. Olav Hellwig (HZDR, TU Chemnitz) for providing access to the XRD tool. 
This research was carried out in part at the Ion Beam Center and ELBE large-scale facilities at the Helmholtz-Zentrum Dresden-Rossendorf e.V., member of the Helmholtz Association. This work is financed in part via the ERC grant 3DmultiFerro (project number: 101141331). 

\bigskip

\section*{Author contributions}
C.O. and D.M. developed the project idea. 
R.B.A. developed the theory of bulk and surface quantum metric quadrupoles with support from M.T.M. and C.O.. G.N. performed the symmetry analysis of nonlinear transport. M.T.M. performed the theory analysis of the nonlinear conductivity in bismuth thin films. 
P.M. prepared samples and performed magnetotransport characterization and electrical harmonic measurements with the support from I.V. and D.M.. 
P.C. carried out electron microscopy measurements. 
I.V. and F.G. performed XRD measurements. 
O.P. performed optical conductivity measurements and analyzed the data. The analysis of the transport data was done by P.M., I.V., M.T.M., D.M. and C.O. with support from R.B.A., G.N., and Deb.M.. 
S.Kov., S.Kob. and Z.W. performed THz harmonic generation studies.
The manuscript was written by R.B.A., P.M., G.N., M.T.M., D.M., and C.O. with the contribution from I.V., Deb.M., P.C., O.P. and F.G.. 
	
\section*{Competing interests}
The authors declare no competing interests.


%

\end{document}